\documentclass[aps,prb,twocolumn,preprintnumbers,superscriptaddress]{revtex4}

\usepackage{graphicx}
\usepackage{dcolumn}
\usepackage{bm}

\begin{document}


\title{Colossal electroresistance and colossal magnetoresistive step in paramagnetic insulating phase of  single crystalline bilayered manganite(La$_{0.4}$Pr$_{0.6}$)$_{1.2}$Sr$_{1.8}$Mn$_{2}$O$_{7}$  }

\author{Y.Yamato}
\author{M.Matsukawa} 
\email{matsukawa@iwate-u.ac.jp }

\author{Y.Murano}
\affiliation{Department of Materials Science and Engineering, Iwate University , Morioka 020-8551 , Japan }
\author{R.Suryanarayanan}
\affiliation{Laboratoire de Physico-Chimie de L'Etat Solide,CNRS,UMR8182
 Universite Paris-Sud, 91405 Orsay,France}
\author{S.Nimori}
\affiliation{National Institute for Materials Science, Tsukuba 305-0047 ,Japan} 
\author{M.Apostu}
\affiliation{Department of Physical, Theoretical and Materials Chemistry, Faculty
of Chemistry, Al. I. Cuza University, Carol I, 700506 Iasi, Romania
}
\author{A.Revcolevschi}
\affiliation{Laboratoire de Physico-Chimie de L'Etat Solide,CNRS,UMR8182
 Universite Paris-Sud, 91405 Orsay,France}
\author{K. Koyama}
\author{ N. Kobayashi}
\affiliation{Institute for Materials Research, Tohoku University, Sendai  
980-8577, Japan}

\date{\today}

\begin{abstract}
We report  a significant decrease in the low-temperature resistance induced by the application of an electric current on the $ab$-plane in the paramagnetic insulating (PMI) state of (La$_{0.4}$Pr$_{0.6}$)$_{1.2}$Sr$_{1.8}$Mn$_{2}$O$_{7}$.  A colossal electroresistance effect attaining $-95 \%$ is observed at lower temperatures.  A colossal magnetoresistive step appears near 5T at low temperatures below 10K, accompanied by an ultrasharp width of the insulator-metal transition. 
Injection of higher currents to the crystal causes a disappearance  of  the steplike transition. These findings have a close relationship with the presence of  the short-range charge-ordered clusters pinned within the PMI matrix of the crystal studied. 
\end{abstract}

\maketitle
%

%
Perovskite manganites show a great variety of fascinating  properties such as colossal magnetoresistance (CMR) effect and charge-ordered (CO) insulating  phase \cite{TO00}. The most interesting one is the existence of a phase-separated state, the coexistence of antiferromagnetic (AFM) charge-ordered insulating   and  ferromagnetic  metal (FMM) regions\cite{DA01}. Several recent works on metamagnetic transitions in phase-separated manganites have revealed that  ultrasharp steps in magnetization curves appear at low temperatures \cite{MAH02,HA03}. The author has proposed that  the magnetization step of manganites arises from a martensitelike transformation associated with lattice strain between competing FMM and CO phases but questions have been  raised \cite{HA03}. 

Recently, it has been reported that the resistance of charge-ordered manganites exhibits a significant change when a strong electric field or current is applied\cite{AS97,RA00}.  
The electric current-induced  resistance change results in a switching from the charge-ordered insulating to ferromagnetic metallic state. This phenomenon is called   a colossal electroresistance (CER) effect, which is analogous to the term of CMR effect. 
The effect of electric fields on the CO state of the manganites is an issue of considerable interest not only on the basis of   physical phenomena  but also from a viewpoint of technological applications. 
There have been extensive studies on the effect of an applied electric current/voltage on the resistance of cubic manganites causing a huge negative ER effect so far \cite{AS97,RA00,ME02,JA06}. However, they have  paid little attention on a bilayered manganite system showing an enhanced CMR effect due to its two-dimensional double layers of MnO$_{2}$ sheets.

In this letter, we report a significant decrease in the low-temperature resistance induced by the application of an electric current on the $ab$-plane in the paramagnetic insulating (PMI) state of Pr-substituted (La$_{0.4}$Pr$_{0.6}$)$_{1.2}$Sr$_{1.8}$Mn$_{2}$O$_{7}$. 
In our previous work, a steplike lattice deformation associated with the ultrasharp PMI-FMM transition has been investigated for single crystalline bilayered manganite (La$_{0.4}$Pr$_{0.6}$)$_{1.2}$Sr$_{1.8}$Mn$_{2}$O$_{7}$\cite{MA07L}.
Next, we try to investigate the effect of an electric current on the colossal magnetoresistive step in the PMI phase of the crystal.   


For the Pr-substituted (La$_{1-z}$,Pr$_{z}$)$_{1.2}$Sr$_{1.8}$Mn$_{2}$O$_{7}$ ($z$=0.6) crystal,  a spontaneous ferromagnetic metal phase (originally present with no Pr substitution) disappears at  ground state but a field-induced PMI to FMM transformation is observed over a wide range of temperatures\cite{MO97,AP01}.
A magnetic  ($H,T$) phase diagram, established from  magnetic measurements, is separated into three regions labeled as PMI, FMM , and bistable states, as shown in Fig.1 of ref.\cite{MA05}. 
Single crystals were grown by the floating zone method using  a mirror furnace. The calculated lattice parameters of the tetragonal crystal structure of the crystals used here were shown in a previous report\cite{AP01}. The dimensions of  the $z$=0.6 sample are 3.5$\times$2 mm$^2$ in the $ab$-plane and 1.7 mm along the $c$-axis. The temperature dependence of resistivity was measured by using a closed cycle refrigerator with a conventional four-probe method. In particular, the electrodes  were  prepared  using a gold paste, to make  a passage of the current uniform. 
Measurements of magnetoresistance(MR)   were done  using a 15-T superconducting magnet at Institute for Materials Research, Tohoku University. The normal sweep rate was set to be 0.26 T/min.
Once we measured  the isothermal MR at selected temperatures, the sample was kept fixed at high temperature of 150 K for 2 hours for demagnetization. Checking the stability of sample's temperature, we restart recording the MR data.
In our field-cooled (FC) measurements, the sample was cooled  from 150 K down to 4 K under the application of a magnetic field, we then removed the applied field and finally  collected the MR data. 

\begin{figure}[ht]
\includegraphics[width=8cm,clip]{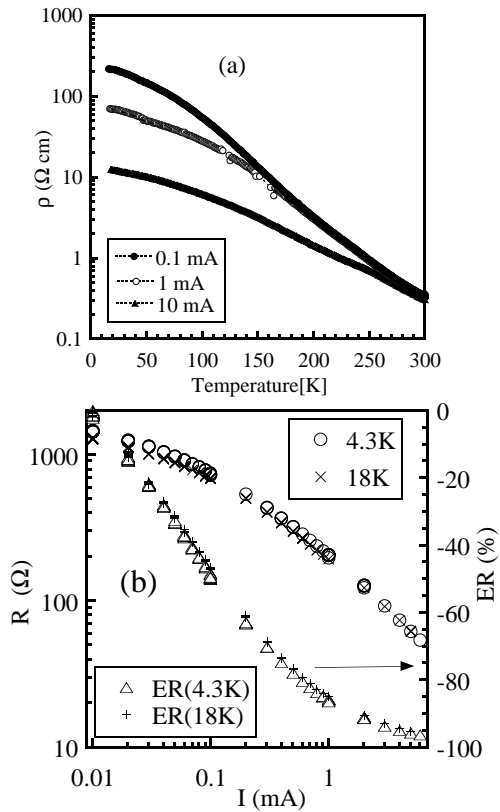}%
\caption{(a)The temperature variation of $ ab$-plane resistance of the  $z$=0.6 crystal measured for various applied currents of 0.1, 1 and 10 mA. (b) The electroresistance(ER) and the ER ratio on the $ ab$ plane at lower temperatures of both 4.3 and 18 K up to 6 mA.}
\end{figure}%
\begin{figure}[ht]
\includegraphics[width=8cm,clip]{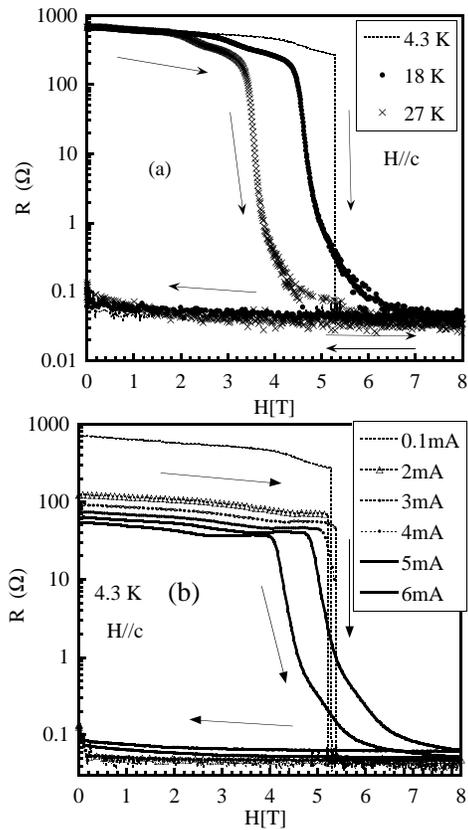}%
\caption{ (a) $ab$-plane magnetoresistance data of  the $z$=0.6 crystal at selected temperatures of 4.3, 18 and 27 K for $I$=0.1mA. ($H||c$) (b) $ab$-plane MR data as a function of an applied current from 0.1 up to 6 mA.}
\end{figure}%
\begin{figure}[ht]
\includegraphics[width=8cm,clip]{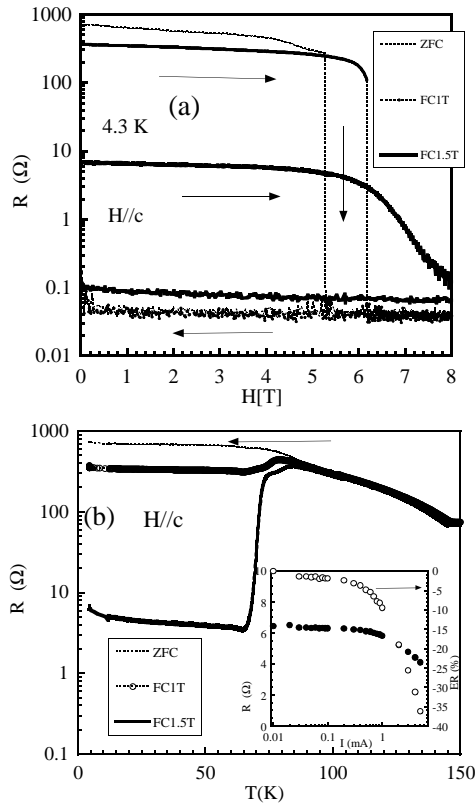}%
\caption{ (a) $ab$-plane MR data of  the $z$=0.6 crystal after the field cooling (FC) runs, FC-1T and FC-1.5 T, at $I$ = 0.1 mA ($H||c$).  For comparison, the zero-field cooling (ZFC) data are presented. (b)  the temperature profile of the resistance monitored in the FC process from 150 K down to 4 K.  The inset  presents the ER and the ER ratio in the FC-1.5 T case at 4.3 K.}
\end{figure}%


Let us show in Fig.1 the temperature variation of ab-plane resistance R$_{ab}$(T) of the z=0.6 crystal measured for several applied currents of 0.1, 1 and 10 mA. 
The substitution of Pr ion for La site suppresses the PMI-FMM transition temperature $T_{c}$= $\sim$ 120K in the case of Pr-free crystal down to $T_{c}$= $\sim$ 90 K at $ z=0.2$.  For the present crystal with $z=0.6 $, a spontaneous transition  disappears even at the lowest temperature of 4 K and the system remains the PMI stable phase. 
At high current of 10 mA, the temperature variation  of $R_{ab}$ tends to saturate in the lower temperature region.  
Next, we examine the effect on an electric current  on the  resistance $R$($I$) of the $z=0.6 $ crystal at the fixed temperatures of 4 and 18 K. The current dependence of $R$ follows a strong nonlinearity  and the magnitude of $R$ rapidly drops by more than one-order of magnitude at higher currents. We note that the rise of the bath temperature  remains at most $\delta T\sim$ 0.2 K even at the higher current of 5 mA.  Joule heating effect is almost negligible because of thermal exchange between the sample and the heat bath. 
Here, an electroresistance ratio  is defined as  $[R(I)-R(I_{low})]/R(I_{low})\times  $100, where $I_{low}$ is  the lowest current value(0.01mA). 
A colossal electroresistance effect reaching ER$=-95 \%$ at $I=5$mA is observed at lower temperatures. It is sure that a lower resistance state is achieved at higher currents but a current-induced insulator-metal (IM) transition observed in the charge-ordered cubic manganites is not noticed up to 10 mA. 
We expect that the lower resistive state is realized by removing scattering centers of charge carriers due to the application of higher currents, such as the charge-ordered clusters being present within the PMI matrix. The formation of short-range charge ordering in the paramagnetic phase of an optimally doped parent crystal La$_{1.2}$Sr$_{1.8}$Mn$_{2}$O$_{7}$ has been observed in a previous study\cite{VA99}. Furthermore, neutron scattering measurements on a bilayered manganite La$_{2-2x}$Sr$_{1+2x}$Mn$_{2}$O$_{7}$ ($x=0.38$) have revealed that the CE-type (zigzag FM chains with AFM interchain coupling) CO clusters freeze upon decreasing temperature from  $T\sim310$K, preventing the formation of a long-range CO state \cite{AR02}. 
Here, we assume that the small CO clusters are pinned to the lattice through Jahn-Teller effect of Mn$^{3+}$ ions as pointed out in ref. \cite{GU99}. We then understand that they are depinned by the application of strong electric field, resulting in the nonlinear ER effect. Moreover, upon increasing $ T$, a thermally activated depinning becomes more dominant, leading to a smaller ER effect\cite{RA00}.  This pinning model qualitatively describes the CER effect enhanced at low $T$. As for the CER effect in the ferromagnetic insulating phase  of La$_{0.82}$Ca$_{0.18}$MnO$_{3}$, the authors argue that the electric current enhances transfer integral between  $e_{g}$ orbitals of neighboring Mn sites i.e., increases the $e_{g}$ bandwidth, which results in a resistance drop \cite{ME02}. To our knowledge, we have no experimental evidence for the enhanced bandwidth effect due to the application of a strong electric field on manganites.

Figure 2 shows magnetoresistance data describing a magnetic-field induced IM transition  at selected temperatures of 4.3, 18 and 27 K ($H||c$). The MR ratio,$[R(H)-R(0)]/R(H)\times  $100, attains $ \sim 10,000 \%$ at 8 T, exhibiting the typical CMR effect. 
A colossal magnetoresistive step suddenly appears with an ultrasharp width ($ \sim $ 10mT)  of the insulator-metal transition when the applied field exceeds  the critical field ($H_{c}\sim 5$T)  at the lowest temperature below 10 K.  On the other hands, the higher temperature MR data follow a broad variation with the transition width ($ \sim $1T).
Now, we present in Fig.2(b) the influence of electric current on the MR at 4.3 K. First of all, the zero-field resistance $R(0)$ is reduced upon increasing the current passing the $z=0.6 $ crystal as shown in Fig.1(b).  Next, at higher currents beyond 4 mA, the magnetoresistive step completely disappears and instead we obtain a broad  IM transition  in magnetoresistance similar to that realized at higher temperatures
The application of higher currents suppresses the steplike variation, resulting in a smooth change in MR. If we term an onset field signifying the field-assisted IM transition as $H_{on}$, the value of $H_{on}$ shifts a lower field from 5 T down to 4 T upon increasing the current from 4 mA up to 6 mA.  
We find out that the steplike phenomena are depressed due to application of higher current. This finding implies a manifestation that  the presence of small CO clusters pinned to the lattice is close to the appearance of the step transition.
 
Finally, we examine the effect of field-cooling on the MR of the $z=0.6 $ crystal since the field-cooling process increases a volume fraction of the FMM phase against the PMI matrix. The $ab$-plane MR data are presented in Fig.3(a) after the sample experienced the FC mode (cooling field 1 and 1.5 T).  A low resistive state ($ \sim 6\Omega $) is realized  in the case of  FC-1.5T  before the MR measurement, as shown in the temperature profile of $R$ of Fig .3(b). 
The FC process causes a substantial suppression of the steplike transition from  $H_{c}\sim 5 $T at ZFC to 6.2 T at FC-1T and for FC-1.5T, we then obtain the standard field-induced IM transition similar to that observed at higher $T$. The ER ratio is reduced to be $=-35 \%$ at $I=$5 mA through FC-1.5T because the crystal is transformed to the low resistive state containing the large FMM region. This ER effect is, in its origin,different from the CER effect in the PMI phase, and is probably   attributed to a structural rearrangement of FMM clusters into filamentary shapes along the direction of an applied current\cite{DO07}. 
In summary, we have demonstrated the current-induced nonlinear decrease in the low-temperature resistance in the PMI phase of bilayered manganite (La$_{0.4}$Pr$_{0.6}$)$_{1.2}$Sr$_{1.8}$Mn$_{2}$O$_{7}$. 
The colossal magnetoresistive step at low temperature is suppressed due to injection of higher currents to the crystal, resulting in a smeared variation in MR. 

This work was supported by a Grant-in-Aid for Scientific Research from Japan Society of the Promotion of Science.


\begin{thebibliography}{30}
\bibitem{TO00} \textit{Colossal Magnetoresistive Oxides}, edited by Y.Tokura (Gordon and Breach,New York,2000).

\bibitem {DA01}\textit{Nanoscale Phase Separation and Colossal Magnetoresistance}, by E.Dagotto(Springer,2003).

\bibitem{MAH02}R.Mahendiran,A.Maignan,S.Hebert,C.Martin,
M.Hervieu,B.Raveau,J.F.Mitchell,and P.Schiffer,
Phys.Rev.Lett.89,286602(2002).
\bibitem{HA03}V.Hardy,A.Maignan,S.Hebert,C.Yaicle,C.Martin,
M.Hervieu,M.R.Lees,G.Rowlands,D.Mc K. Paul, and B.Raveau,
Phys.Rev.B68,220402(R)(2003).
\bibitem {AS97}A. Asamitsu, Y. Tomioka, H. Kuwahara, and Y. Tokura, 
Nature(London) 388, 50(1997).
\bibitem {RA00}C.N.R.Rao,A.R.Raju,V.Ponnambalam,Sachin Parashar,N.Kumar,Phys.Rev.B61,594(2000).
\bibitem{ME02}S.Mercone,A. Wahl,Ch. Simon, and C. Martin,
Phys.Rev.B65,214428(2002).
\bibitem{JA06}Himanshu Jain,A.K.Raychaudhuri,Ya.M.Mukovski,D.Shulyatev,
Appl.Phys.Lett.89,152116(2006).
\bibitem{MA07L}M.Matsukawa,Y.Yamato,T.Kumagai,A.Tamura, 
R.Suryanarayanan, S.Nimori, 
 M.Apostu, A.Revcolevschi, K. Koyama and N.Kobayashi, 
Phys.Rev.Lett. 98,267204(2007).
\bibitem{MO97}Y.Moritomo, Y.Maruyama,T.Akimoto, and A.Nakamura, 
Phys.Rev.B56,R7057(1997).
\bibitem{AP01}M.Apostu, R.Suryanarayanan, A.Revcolevschi, 
H.Ogasawara, M.Matsukawa, M.Yoshizawa, 
and N.Kobayashi,Phys.Rev.B64,012407(2001).
\bibitem{MA05}M.Matsukawa,K.Akasaka,H.Noto, R.Suryanarayanan, S.Nimori, 
M.Apostu, A.Revcolevschi, and N.Kobayashi, 
Phys.Rev.B72,064412(2005).
\bibitem{VA99}L.Vasiliu-Doloc, S. Rosenkranz,R. Osborn,
S. K. Sinha,J. W. Lynn,J. Mesot,O. H. Seeck,G. Preosti,
A. J. Fedro,and J. F. Mitchell,
Phys.Rev.Lett.83,4393(1999).
\bibitem{AR02}D.N.Argyriou,J.W.Lynn,R.Osborn,B.Campbell,
J.F.Mitchell,U.Ruett,H.N.Bordallo,A.Wildes,and C.D.Ling,
Phys.Rev.Lett.89,036401(2002).
\bibitem{GU99}A. Guha, A. Ghosh, A.K.Raychaudhuri,S.Parashar,
A.R.Raju,C.N.R.Rao 
Appl.Phys.Lett.89,152116(2006).
\bibitem{DO07}S.Dong,H.Zhu,J.M.Liu,Phys.Rev.B76,132409(2007).

\end{thebibliography}
\end{document}